# An Adaptive Networks Model to Simulate Consensus Formation Driven by Social Identity Recognition

Kaiqi Zhang[1,3], Zinan Lv[4], Haifeng Du[2,3,*] , Honghui Zou[4]

[1] *School of Economy and Management, Chang'an University, Xi'an, China*

[2] *School of Public Policy and Management, Xi'an Jiaotong University, Xi'an, China*

[3] *Center for Public Management and Complexity Science Research, Xi'an Jiaotong University, Xi'an, China*

[4] *School of Automobile, Chang'an University, Xi'an, China*

Correspondence should be addressed to Kaiqi Zhang: kaiqizhang@chd.edu.cn

**Abstract**

Models of the consensus of the individual state in social systems have been the subject of recent researches in the physics literature. We investigate how network structures coevolve with the individual state under the framework of social identity theory. And we propose an adaptive network model to achieve state consensus or local structural adjustment of individuals by evaluating the homogeneity among them. Specifically, the similarity threshold significantly affects the evolution of the network with different initial conditions, and thus there emerges obvious community structure and polarization. More importantly, there exists a critical point of phase transition, at which the network may evolve into a significant community structure and state-consistent group.

## 1．Introduction

With the development of Internet, the method of information spreading mainly depends on online social media, which promotes the formation of consensus in the complex network[1-4]. Understanding and analyzing the consensus formation can help the explanation and governance of social phenomena, such as public state polarization[5], rumor spreading[6] or riots, segregation, and interactions in cities [7]. The latest research on complex networks shows that consensus is the result of the coevolution of network structure and individual state. And the formation process of consensus can be revealed by an adaptive network[8,9].

An adaptive network model is a dynamically changing model that can self-organize network connections and update individual states. Models of the consensus in social systems have been the subject of a considerable amount of recent researches in the physics literature[10]. These models can be categorized into two classes. One focuses on the role of network structure or individual state in the dynamic evolution process. Holme et al.[10,11] (hereinafter referred to as the Holme model) proposed a model based on the well-known Voter model[12,13]. Their model combines state dynamics with assortative network formation, revealing an apparent phase transition between regimes in which one process or the other dominates the dynamics[10]. The other tries to grasp the dynamic interaction of the individual state.

A popular model was proposed by Kozma et al.[4] (hereinafter referred to as the Kozma model), who investigated how the coevolution of an adaptive network of interacting individuals and the individual's states influence each other, and how the final state of the system depends on this coevolution. Their model is based on the Deffuant model[14] for which a large number of states can coexist (and not only 2 as in the Voter model)[4].

It is observed in these models that network structure, coevolution rule, and distribution of state among social individuals significantly affect the final structure of the network and distribution of state. Furthermore, we find that these models undergo a continuous phase transition, from a regime in which the state is arbitrarily diverse to one in which most individuals hold the same state. And it can be controlled by adjusting the relevant parameters of the model [2,3,12,15,16].

It can be found that most models try to reveal a general law of consensus by constructing an adaptive network. state convergence can be observed in social media where social networks tend to divide into several groups or communities of individuals with similar states. And adaptive rewiring of links can happen in real-life systems such as acquaintance networks where people are more likely to maintain a social connection if their views and values are similar. There is an obvious question to ask is why individuals want to change their states or rewiring their social connections. In another phrase: what are their initial needs and underlying motivations to change state or adjust structure? However, the Holme model can not reflect the mutual interaction between individuals in social media. It requires prior knowledge, which depends on certain network structure and individual state information, as a part of coevolution rules. Improved models like the Kozma model focus on simple evolution rules that do not require prior knowledge of the states of individuals to which new links are established. But in their models, all individuals have the same behavior choices in response to different group effects. In other words, individuals are homogeneous, and their heterogeneity is not well-reflected.

Up to now, few studies however considered the fact that individuals have different responses when they are in different groups, or even they are in the same group they may still make different choices because they are heterogeneous. In addition, in the individual's cognition of the message they received, the "herding effect" is particularly considered. It is believed that individuals will not only make judgments based on their rational expectations but also make decisions based on public behavior. Therefore, we propose a model that focuses on the realistic needs of individuals in a specific environment and the impact of behavioral motivation. To better understand the formation of consensus, social norms, and individual psychology are considered in our model. Specifically, we introduce the social identity theory explaining social comparison and self-categorization when individuals face behavior choice. Social comparison and self-categorization both can give rise to realistic levels of agreement between acquaintances. Here we combine both processes with a single parameter---*similarity threshold*---controlling the balance of the two processes. We investigate the role of the various parameters such as the modularity and the rate of the largest group. We show that the network structure and evolution rules have important consequences on the evolution mechanisms of consensus formation.

The contributions of our work are threefold: 1) There's only one free parameter (similarity threshold), which can highlight the heterogeneity of individuals in the coevolution, in our model. This parameter has a clear meaning in social governance, which is a practice like disclosing and dispelling false information on social media, policy guidance and behavioral intervention; 2) We make coevolution rules based on social identity theory, individuals' behavior choices are highly influenced by local relationships, and majority state in a group has a significant effect on individuals; 3) Based on the BA network, we use similarity threshold to explain the polarization and formation of states in social media.

The organization of the paper is as follows: In Section 2, the relevant issues about the basic rules of the existing adaptive network model and social identity theory are discussed. To give a full description of the model, we start our study with the improved evolution rule and methods in Section 3. Then, we discuss the evolution process and experiment results in Section 4. Finally, in Section 5, there are conclusions of this paper.

## 2．Related Work

### 2.1 Adaptive Network Model Based on Structural Coevolution

The adaptive network addresses the self-organization of complex network structure and its implications for system behavior [17,18,19]. Over the past decades, adaptive networks have been developed and applied to various subjects, ranging from physical, biological to social, and engineered systems. Applications of adaptive networks also include the evolution of an organizational network, information/knowledge/culture sharing, and trust formation within a group or corporation[20,21]. Our study of adaptive networks is in the field of consensus of which models proposed by Holme et al.[10] and Zanette et al. [22] are seminal works.

According to the setting of the Holme model, consider a network of $N$ nodes, representing individuals, joined in pairs by $M$ links, representing social connections between individuals. Each individual is assumed to hold one of $K$ possible states on some topic of interest. The state of individual $i$ is denoted as $k_i$. In each iteration of the model simulation, nodes will perform two basic operations according to the rules: 1) The links are randomly selected and placed between nodes with the same or similar state, or 2) The node changes its state to be consistent with the state of the surrounding nodes. Specific rules are shown in FIG.1:

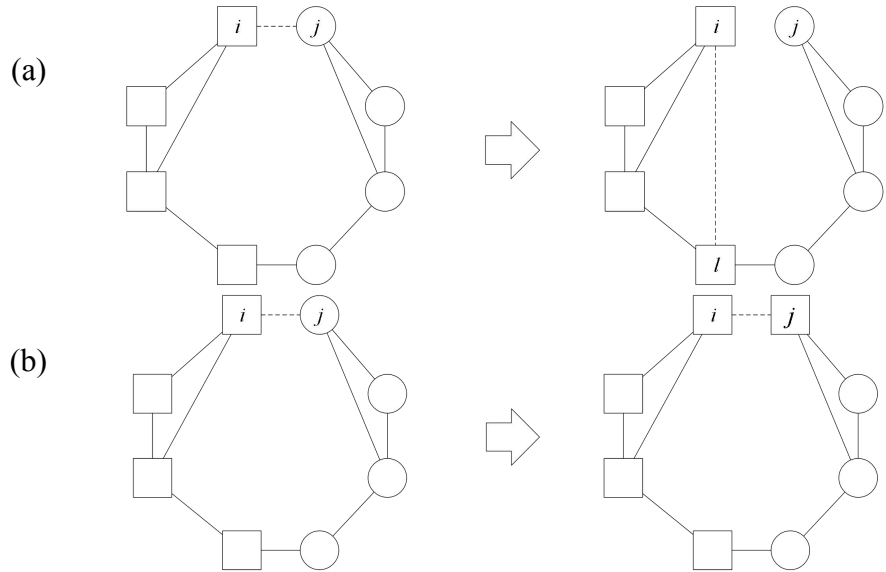

FIG. 1 An illustration of the Holme model, with node shapes representing states. At each time step, the system is updated according to the process illustrated in (a) with probability $\xi$ or (b) with probability 1- $\xi$. In (a) a node $i$ is selected at random and one of its links (in this case the link $(i, j)$) is rewired to a new node $l$ holding the same state as $i$. In (b) node $i$ affects one of its neighbor $j$ and makes $j$ adopt the state of it.

Step (a) represents the formation of new social connections between people of similar states. Step (b) represents the impact of acquaintances on one another. However, the rewiring operation in Step (a) requires certain prior knowledge to guide it. Meanwhile, the one-way influence operation in Step (b) cannot reflect the process of interaction between nodes. Therefore, Kozma et al. [4] proposed an adaptive network model based on the continuous distribution view.

Based on the Holme model and affected by the Deffuant model[12], the Kozma model is proposed. At each time step $t$, two neighboring nodes are selected, and they communicate if their states are close enough, i.e., if $|o(i,t) - o(j,t)| \leq d$, where $d$ defines the tolerance threshold. In this case, the local communication tends to bring states even closer, according to Eq.(1):

$$o(i,t+1) = o(i,t) - \mu[o(j,t) - o(i,t)]$$
$$o(j,t+1) = o(j,t) + \mu[o(j,t) - o(i,t)]$$
(1)

Or, an attempt to break the connection between $i$ and $j$ is made: if $|o(i,t) - o(j,t)| > d$, a new node $l$ is chosen at random and the link $(i, j)$ is rewired to $(i, l)$. The tolerance threshold $d$ has an important impact on network regulation, individuals states tend to be homogeneous as $d$ increases. Compared to the Holme model, the Kozma model emphasizes the micro-interaction between nodes and controls the state convergence through tolerance threshold $d$. Also, the Kozma model emphasizes the randomness in the evolution to reduce the model's dependence on prior knowledge.

It can be found that interaction rule between nodes is the core part of model construction, and the interaction rules in all existing models contain both adjustments of node state and structure. Such adjustment of nodes would affect the whole network structure. However, in most models, the adjustment of one node is affected by another node, and the effect of local structure to the node is completely ignored. Taking the node $j$ in Fig. 1(b) as an example, $j$ is consistent with the state of node $i$ with probability 1- $\xi$ in the Holme model. But according to the social norm, group state have a great impact on individuals. So, the impact of $i$ on $j$ may be less than that of the other two neighbor nodes. With such local structure, abstracting the behavior choice of $j$ only based on the probability 1- $\xi$ does not conform to reality. In the same case, in the Kozma model, the interaction between $i$ and $j$ can be influenced by other neighbor nodes. Moreover, the heterogeneity of nodes is also ignored. In real life, every individual is independent, and his/her behavior choice is affected by various factors such as personality, willpower, education, etc. But the rewiring probability $\xi$ in the Holme model or the tolerance threshold $d$ in the Kozma model applies to all nodes, which means all nodes in their models are homogeneous. Therefore, existing models are not very similar to what we may observe in real life. The heterogeneity of individuals should be considered, and the interaction between individuals and groups affected by social norms needs to be designed with corresponding rules. So related theories in sociology need to be introduced to improve related rules.

**2.2 The Formation of consensus in Social Identity Theory**

Social identity theory stems from the description of group formation and group relationship by ethnocentrism and realistic conflict theory[23]. Tajfel and Turner proposed social identity theory in 1986 to theorize how people conceptualize themselves in intergroup contexts and how a system of self-categorizations "creates and defines an individual's place in society" [24,25]. Tajfel defined *social identity* as "the individual's knowledge that he belongs to certain social groups together with some emotional and value significance to him of this group membership". Social identity can not only help individuals to identify the same social attributes, integrate into specific social groups, but also enable individuals to confirm their value realization. This value realization includes both the value of physical capital and the satisfaction of psychological cognition. FIG.2 shows the two stages of social identity: self-categorization and social comparison.

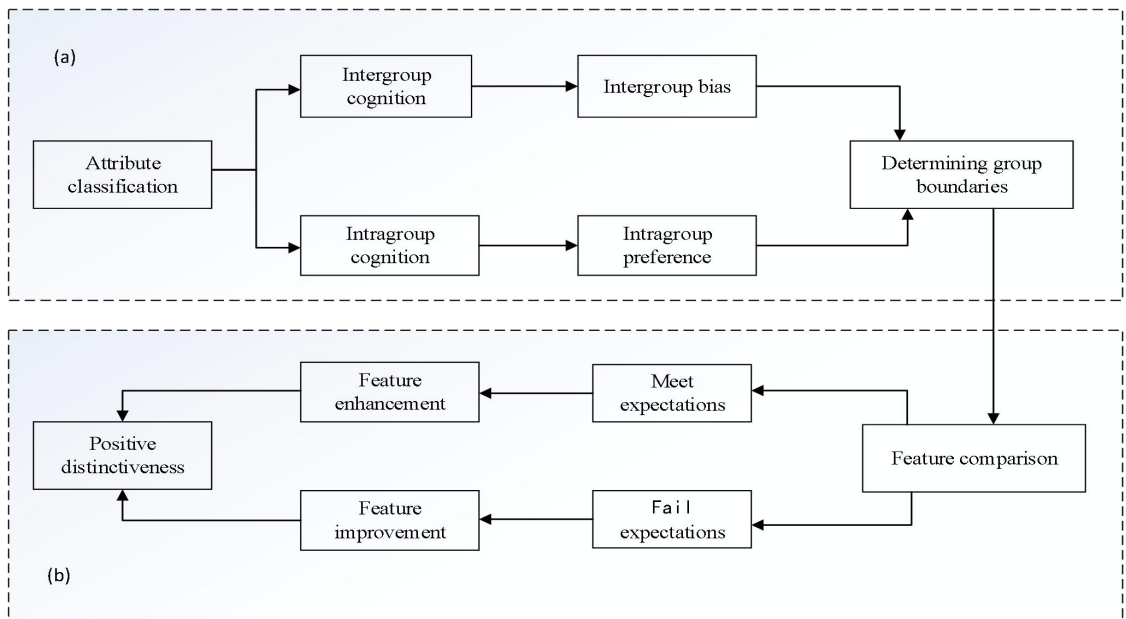

FIG.2 Formation of social identity. Panel(a) shows the self-categorization process. Everyone can be subordinated to one or more social groups according to different attributes. And each group has the same attribute. At the same time, to distinguish the subordinate groups from other groups, an individual's cognition of the in-group similarity will continue to increase, and the out-group differences will continue to expand, thus forming obvious attributes and behavioral characteristics belong to a certain group. Panel(b) shows the social comparison process. It focuses on establishing positive heterogeneity for one's group. Positive heterogeneity is a precondition for the formation of social identity.

Self-categorization simplifies an individual's cognition of human society. Individuals usually use simple and independent attributes such as ethnicity, race, class, occupation, gender, etc. to describe things. They categorize themselves into a group with the same or similar attribute to reduce cognition of the complex world. Different attributes and state disagreements create group boundaries. And the group boundary formed in self-categorization is the precondition of the feature comparison which triggers the process of social comparison. The differences between groups are magnified in social comparison, thus form positive heterogeneity. Pursuing positive heterogeneity is also a key part for individuals to realize their value and complete the evolution from individual behavior to consensus.

Overall, social identity enhances individual cognition and the positive heterogeneity formed in social identity is also the internal psychological driving force of individuals' tendency to group behavior. In the interaction between individuals and groups, the differences between individuals will gradually be weakened through the formation of social identity, which will also help to modify individual behavior choices, thereby forming a consensus. But existing models lack interaction rules between individuals and their groups, so we introduce social identity theory as an interaction rule of individual behavior. We design specific rules based on self-categorization and social comparison to better understand the internal mechanism of consensus.

## 3. The Proposed Model

In this section, we propose an adaptive network model describing the formation of consensus, which is based on social identity theory.

### 3.1 Model Definition

According to the above description of the classical social identity theory, social identity lead to consensus. There, we incorporate the social norm into the model rules and defines it as a social state of individuals with similar states or attributes. Here we set an *r*-dimensional vector abstracting the individual multiple states or attributes and represent it as $\omega_i = \{a_1^i \, a_2^i \ldots a_r^i\}$, where $a_k^i$ represents the *k*th state of the individual *i*. The state difference between individuals can be expressed by the Euclidean distance

$$d(i,j) = \sqrt{\sum_{k}^{r} \left(a_k^i - a_k^j\right)^2} \tag{2}$$

To normalize Eq.(2) $d(i,j)$, there are

$$sim_{i,j} = \frac{1}{1 + d(i,j)} \tag{3}$$

$sim_{i,j}$ represents the similarity between *i* and *j*. Formally, $sim_{i,j}$ close to 1 indicates a higher similarity between individuals, otherwise, it indicates a lower similarity. From the perspective of social identity theory, the self-categorization of social individuals needs to judge the similarity between individuals according to specific social standards. A similarity threshold is introduced here to help with the self-categorization of individuals. The function of self-categorization is given by

$$\delta(i,j) = \begin{cases} 1 & sim_{i,j} \geq \theta \\ 0 & sim_{i,j} < \theta \end{cases} \tag{4}$$

where $\theta$ is the similarity threshold. When $sim_{i,j} \geq \theta$, it is considered that individuals are homogeneous, and if there is a social interaction between *j* and *i*, it is determined that *j* is a homogeneous neighbor individual of *i*, otherwise, it means that there is heterogeneity among individuals. Like the rewiring mechanism in the Holme model and the Kozma model, we abstract self-categorization as a relationship modification process. The establishment of homogeneous relationships is the goal of each individual. And we call this process the Structural Update Mechanism (SU mechanism).

While updating the structure of social relations, strengthening individuals' cognition is an important part in the formation of consensus. Inspired by the Kozma model's interaction rules, we design a Collective Adaptive Mechanism (CA mechanism) to abstract the strengthening of individuals' cognition. Specifically, it is assumed that set $T_i$ represents a set of neighbor individuals of the *i*. According to the Eq.(4), $T_i$ can be subdivided into a set of homogeneous individuals $s_i = \{m \mid \delta(i,m)=1,\}$ and a set of heterogeneous individuals $h_i = \{n \mid \delta(i,n)=0\}$ ($s_i \cup h_i = T_i$). In particular, individuals only update their states with neighbors in $s_i$, but not in $h_i$. In addition, the individual will update states according to the local average state of $s_i$, the local average state is given by

$$\overline{\omega_i} = \left\{\overline{a_1}, \overline{a_2} \cdots \overline{a_r}\right\}; \overline{a_k} = \frac{\sum_{m \in s_i} a_k^m}{\|s_i\|} \quad (5)$$

where, $\overline{a_k}$ is the average state of the $k$th state. Similar to the interaction rules in the Kozma model, individuals update their states one by one and update $a_k^i$ to a new state $a_k^{i}{'}$

$$a_k^{i}{'} = a_k^i + \mu\left[\overline{a_k} - a_k^i\right] \quad (6)$$

where, $\mu$ is the update parameter of the state, usually $\mu \in (0, 0.5]$. Eq.(6) shows that individual will update its states to make sure the convergence to the average state of its homogeneous neighbors.

The above two mechanisms restore the description of the self-categorization and social comparison process, and present individual cognitive changes in the formation of consensus. Both mechanisms are influenced by the similarity threshold $\theta$. In social media, the greater the similarity threshold $\theta$ is, the more heterogeneous relationships among individuals, and they are more likely to adjust existing structures, such as unfollowing those who disagree with them. While the smaller the similarity threshold $\theta$ is, the more homogeneous relationships between individuals, and they are more likely to strengthen the individual's convergence to the average state.

Compared to the mechanism in models like the Kozma model, there are three differences in our mechanism. Firstly, we introduce the similarity threshold in the SU mechanism, where a local structure is updated by the homogeneity between individuals. It is more in line with the interaction of individuals in the real environment. Secondly, in the CA mechanism, we do not introduce relevant parameters such as the tolerance threshold $d$ which has been talked in the Kozma model. Instead, the individual's cognitive update is realized by the average state of homogeneous neighbors, emphasizing the interaction between individuals and groups, reflecting the individual's convergence in the formation of consensus. Thirdly, different from the obedience principles in existing models, the convergence in our mechanism emphasizes the impact of local homogeneous individuals. And update rule about the heterogeneous relationships between individuals highlights the impact of social interaction on individual behavior choice. In summary, the mechanism proposed in this paper is derived from the general description of social identity theory. It abstracts the process of individuals classify themselves into a group by different states, and achieving consensus by defining group boundaries. With this process, we can analyze the possible behavioral strategies of individuals based on the specificity of consensus among different groups.

**3.2 Construction of Adaptive Network Model**

Based on the above theoretical modeling, we construct an adaptive network model to simulate the coevolution of the network based on individuals' structure and states. Specifically, set network $G=(V, L)$ represents social relationships between individuals, where $V$ represents a set of nodes formed by $X$ individuals, and $L$ represents a set of connected links formed by $Y$ social relationships. And the state of each node is a set of $r$-dimensional vectors $\omega$, each state of the node $a_k^i$ obeys a continuous distribution within a certain interval. There is only one controllable parameter, which is the similarity threshold $\theta$. Referring to the relevant settings of the existing model, the state update parameter $\mu$ in Eq.(6) is set to a certain value. The update process of the model at time $t$ is divided into three steps as shown in FIG. 3:

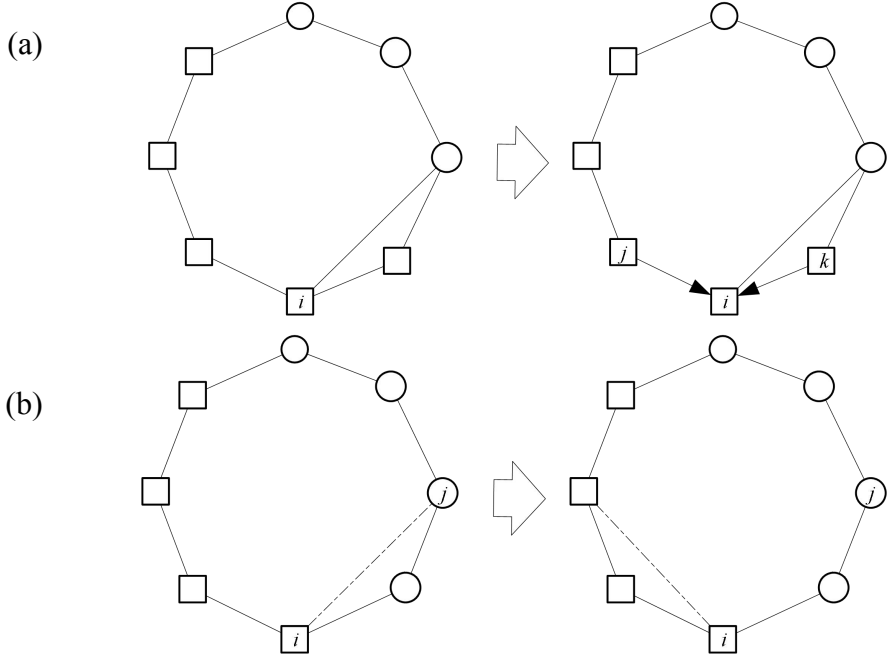

FIG.3 Illustrates SU and CA mechanism at time *t*, with node shapes representing states. At each time step, the system is updated according to the process illustrated in panel (a) or panel (b) with a precondition of the number of homogeneous (heterogeneous) neighbor nodes. When the number of homogeneous neighbor nodes of $v_i$ is greater than that of heterogeneous neighbor nodes, as shown in (a), the state of $v_i$ will be affected by its two homogeneous neighbor nodes. The direction of the arrow indicates the direction in which neighbor nodes influence it. Otherwise, as shown in (b), $v_i$ will randomly select a node, and achieve a rewiring process.

**Step1:** Randomly selecting the node $v_i$, calculating the similarity $s_i$ between $v_i$ and its neighbors based on Eq.(3). At the same time, referring to the similarity threshold $\theta$, the neighbor nodes of $v_i$ are divided into the homogeneous node-set $s_i$ and the heterogeneous node-set $h_i$.

**Step2:** When $\|s_i\| \geq \|h_i\|$, as shown in Fig. 3(a), the number of homogeneous neighbor nodes of $v_i$ is greater than that of heterogeneous neighbor nodes. The homogeneous neighbor nodes of $v_i$ will calculate their average state based on equation (5), here, it is the average state of $v_j$ and $v_k$. Node $v_j$ and $v_k$ will influence the state of $v_i$ in the direction of the arrow. Node $v_i$ updates its state based on Eq.(6).

**Step3:** When $\|s_i\| \leq \|h_i\|$, as shown in Fig. 3(b), the number of homogeneous neighbor nodes of $v_i$ is smaller than that of heterogeneous neighbor nodes. It indicates that $v_i$ is outside the group at the moment. According to the similarity $s_i$ between nodes, use the roulette method to randomly select a node, and achieve a rewiring process. It should be pointed out that the target node of the rewiring link is randomly selected.

Based on the above steps, the overall implementation of the model is shown in Algorithm 1:

| **Algorithm 1 Simulated social identity algorithm** |
|---|
| Input： Network: $G=(V,L)$, Node Status:$\{\omega_i\}$, Number of Iteration Times: $it_{max}$; Similarity Threshold: $\theta$ |
| Output：Network: $G'=(V,L)$, Node Status:$\{\omega_i\}'$ |
| 1： While $t < it_{max}$ |
| 2： For each node $u$ in $V$: |

| | | |
|---|---|---|
| 3： | | Obtain the neighbor nodes set $T$ |
| 4： | | For each node $v$ in $T$: |
| 5： | | Calculate the similarity between node u and node v: $sim_{uv}$; |
| 6： | | If $sim_{ij} \geq \theta$ : |
| 7： | | $s_u \leftarrow$ node $v$; |
| 8： | | Else: |
| 9： | | $h_u \leftarrow$ node $v$; |
| 10： | | End For |
| 11： | | If $\| s_u \| \geq \| h_u \|$ : |
| 12： | | *Do Structural Update Mechanism* |
| 13： | | For each $a_k^{mean} \leftarrow \Sigma a_k^n / \| s_u \|$ , $n \in s_u$ ; $a_k^n \in \omega_n$ |
| 14： | | For each $a_k^{u'} \leftarrow a_k^u + \mu[a_k - a_k^u]$ |
| 15： | | Else: |
| 16： | | *Do Collective Adaptive Mechanism* |
| 17： | | Random Select node $n \in h_u$ by Roulette |
| 18： | | Rewiring link $l(u,n) \rightarrow l(u,l)$ |
| 19： | | End If |
| 20： | End For | |
| 21： | $t$++ | |
| 22： | End While | |

Combined with the information dissemination process in social media, we can think that the impact operation in step 2 is actually the forwarding and dissemination of information in social media, and the rewiring operation in step 3 is the shielding and attention of information.

## 4．Experiments and Results

In this section, we analyze the evolution process and results of the model based on a computer simulation experiment and discuss the impact of parameters.

### 4.1 Experiment settings and parameters

The simulation experiment is divided into two parts: 1) Structure evolution experiments of the generated network model, focusing on the impact of parameters on structure evolution under different network structures; 2) State distribution evolution experiments of the generated network model, focusing on the impact of parameters on the distribution of node states under different network structures. For each set of experiments, there are two different initial networks, which are the ER network and the BA network. The reason we use the BA network as an experimental benchmark network is that the relationship structure of most social media such as Twitter Facebook is a scale-free network[1,26]. Specifically, the parameters of the generated network model are shown in TAB.1:

|  | ER network | BA network |
| --- | --- | --- |
| Node size | $N$=500 | $N$=500 |
| Edge size | $M$=1250 | $M$=1491 |
| Model parameter | $p_c$=0.01 | $m_i$=3 |

TAB.1 Parameters of the generated network model. The $p_c$ in the ER network represents the probability of connecting links in the network. $M$ is the average link size of multiple experiments. In the BA network, $m_i$ represents the number of connected links of the new node.

Based on the generated network, the experiment randomly assigns the initial state of nodes. And then simulation experiments are carried out based on the model parameters. The initial parameters are shown in TAB.2:

| Parameters | Value |
| --- | --- |
| Node states: $a_k^i$ | $U[1,10]$ |
| Node states dimension: $r$ | 4 |
| Model iterations: $it$ | 1000 |
| Update parameter of state: $\mu$ | 0.5 |

TAB.2 The initial parameters of the model. To highlight the difference between node states, the value of the node state $a_k^i$ is set to a uniform continuous distribution in the range of [1, 10]. At the same time, the dimension of the node state is limited to 4, to effectively distinguish states between nodes. The time complexity of the model implementation is affected by the size of the link $M$. Therefore, the number of iterations is set to 1000 to ensure the effect of simulation. Also, update parameters of state $\mu$ are set to 0.5 along with the existing model's setting.

The similarity threshold $\theta$ is the only controllable parameter in the model. We focus on it to discuss the simulation results. Specifically, let the similarity threshold change in the range of [0.00, 1.00] in the step of 0.05. According to the description in social identity theory, the formation of consensus is also accompanied by the formation of groups, and there are obvious boundaries between groups. To distinguish the group boundaries in network structure and node state, our experiments will measure the groups that may generate from the structure and state respectively. For the structure groups, it is expressed as community structure[27]. And modularity[28] is used to measure the community structure, the calculation is given by

$$Q = \frac{1}{2Y} \sum_{ij} \left( a_{ij} - \frac{g_i g_j}{2Y} \right) \sigma \left( C_i, C_j \right) \qquad (7)$$

where $a_{ij}$ is adjacency relations between $v_i$ and $v_j$; $g_i$ and $g_j$ represent the degree of the two nodes; $Y$ is the number of edges in the network; and $\sigma(C_i, C_j)$ indicates whether node $v_i$ and $v_j$ belong to the same community. If so, $\sigma(C_i, C_j)$=1, otherwise $\sigma(C_i, C_j)$=0. According to Newman's description of modularity, when $Q \geq 0.3$, it is assumed that the network has a community structure. There are many methods to detect community structure. In this paper, we use modularity $Q$ which proposed in [23] to detect community structure. It can detect the community structure based on the local attributes of nodes, which is suitable for the design in our rules about node states and local structures.

For the state groups, we distinguish them by the Euclidean distance $d_0^i$ between its states' point and the coordinate origin based on the Eq.(2). By statistical sorting of $d_0^i$, a set of nodes within one standard deviation is regarded as a group. And the characteristics of state group is described by the ratio of the largest group's node size $S_{max}$ to the network node size $N$. Formally, the $S_{max}/N$ is closer to 1 indicates that there is only one unique state group in the network; otherwise, there may be multiple state groups.

The experimental environment is Python3.7, and the operating environment is Intel(R) Core(TM) i7-4790 CPU @ 3.60GHz RAM 24.0 GB. Since the model contains random distribution parameters, all the results are the average results of 30 runs under the same parameter setting.

### 4.2 Experiment Results of Structure Evolution

#### 4.2.1 Analysis of the general evolution process

We conduct experiments on the modularity of the network during model evolution. FIG.4 shows the modularity changes during the evolution of the networks.

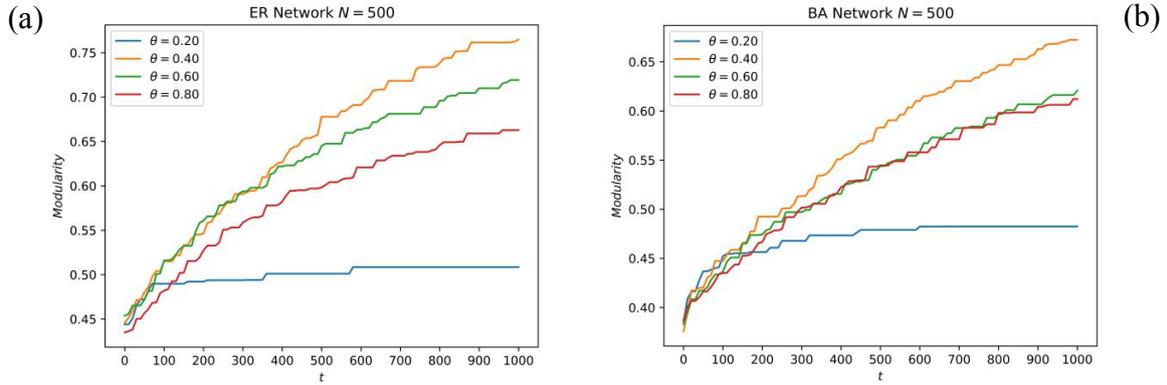

FIG. 4 Evolution results of two different networks under different similarity thresholds.

In general, with the iteration in our model, the modularity of the two networks continues to increase. It indicates that our model can realize the evolution of the network structure and promote the formation of community structure, which is a structure group. Specifically, the SU mechanism in our model can effectively adjust the local structure of nodes, thereby continuously strengthening the community structure and realizing the structure aggregation of individuals. It can be found that the modularity of the ER network is bigger than that of the BA network. We think it is related to the initial network structure. Although the modularity values differ under different similarity thresholds, the evolution law of the model is relatively consistent.

#### 4.2.2 Impact of similarity threshold on structure evolution

We further investigate the impact of the similarity threshold on the structure evolution.

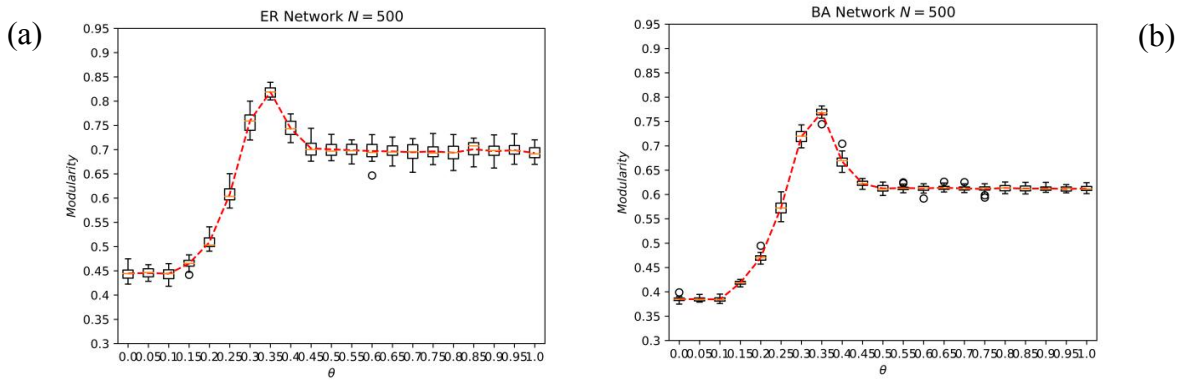

FIG. 5 Illustrates the modularity of the final network structure under different similarity thresholds. (a) and (b) respectively show the modularity of the ER network and the BA network.

As shown in FIG.5, the modularity value is generally affected by similarity threshold $\theta$, and there exists an obvious phase change characteristic. For the two networks, when the similarity threshold is around 0.35, the community structure of the network is most obvious, and the structure group is relatively prominent. In general, the modularity of the evolved networks exhibit evolution characteristic which increases at first then falls back and tends to be stable as similarity threshold $\theta$ increases. At the same time, $\theta=0.35$ and $\theta=0.45$ are two obvious phase transition points. From the distribution of box plots, under the same similarity threshold, the modularity fluctuation of the ER network after evolution is significantly larger than that of the BA network. It indicates that the network degree distribution may have a certain impact on evolution.

It can be found that the similarity threshold has a significant impact on the evolution of the structure group. The modularity phase change is concentrated at the interval of [0.15, 0.45]. And when the similarity threshold is outside the interval, the modularity is stable. It indicates that random initial network structure and random initial node state have less impact on the structure group of the evolved network. In other words, via the model, the final structure of the network can be well controlled and will form a stable structure group. Combining Eq.(4) which is the definition of similarity threshold, similarity threshold controls the number of homogeneous nodes and heterogeneous nodes in the local structure. The larger the similarity threshold, the more heterogeneous nodes in the local structure, and thus the node is more likely to change its current structure. Otherwise, the node has more homogeneous nodes and it is easier to form the convergence of state. However, from the results in FIG.5, the larger similarity threshold does not lead to a more obvious community structure or structure group. On the contrary, that situation occurs when the similarity threshold is more moderate (i.e., around $\theta=0.35$ in the experiment). From the perspective of social identity theory, the similarity threshold can be understood as the evaluation criteria of individual differences in society. The higher the evaluation criteria (the greater the similarity threshold), the greater the social distance between individuals, and individuals are more likely to form stable social relationships and social structures. The lower the evaluation criteria, the smaller the social distance between individuals, and individuals are also likely to form stable social relationships. It can be found that there may be a phase change threshold in the evaluation criteria. When the threshold is exceeded, individuals' need for consensus may increase rapidly, they will form independent groups in structure, which may trigger consensus.

### 4.2.3 Impact of similarity threshold on network degree distribution

We further investigate the impact of the similarity threshold on the degree distribution of the BA network.

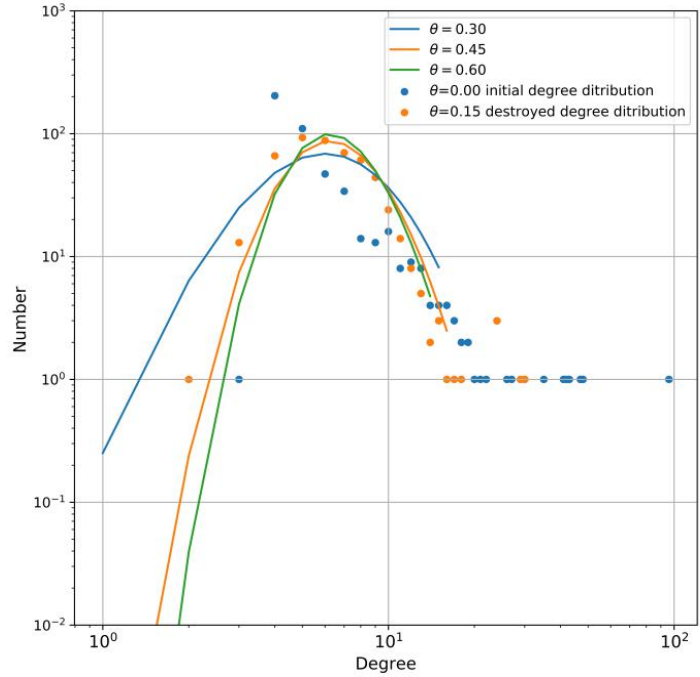

Fig.6 Degree distribution of the BA network in different similarity thresholds. We conducted many experiments on the BA network and obtained its degree distribution. We mainly selected the evolution results of 1000 iterations when $\theta$ is equal to 0, 0.15, 0.3, 0.45, and 0.6. When $\theta$ is greater than 0.6, the evolution result is similar to the result of $\theta=0.6$. The blue scatterplot shows the degree distribution when $\theta=0$, which is the degree distribution of the initial network. The green scatterplot is the degree distribution when $\theta=0.15$. A scatterplot indicates that it follows a power-law distribution. When $\theta>0.15$, the original distribution is destroyed and gradually becomes a lognormal distribution, which we use a curve to represent it. And the blue, orange, and green curves respectively represent the degree distribution when $\theta$ is equal to 0.3, 0.45, and 0.6.

It can be seen in FIG.6 that without the intervention of the similarity threshold, the degree distribution of the BA network shows a power-law distribution. Under the logarithmic coordinate system, the degree distribution of the BA network will eventually show a lognormal distribution. When the similarity threshold is small ($\theta<0.15$), the degree distribution is close to power-law distribution. As the similarity threshold increases to the two phase transition points($\theta=0.15$ and $\theta=0.35$), the Power-law structure is broken and gradually forms a lognormal distribution; With the similarity threshold exceeds 0.35, the structure of the degree distribution tends to be stable.

From the perspective of social identity theory, everyone has equal opportunities to reach consensus with others, thus forming a group. The similarity threshold can be understood as a social barrier. When the similarity threshold is low, social barriers are relatively low, as long as there is a communication, people can reach a consensus, thus forming a whole. But the prerequisite is that people have equal opportunities to communicate. Therefore, it is difficult for them to form large groups, and they can only exist in society as a large number of independent small groups. This phenomenon appears in the area where we can find that the degree distribution of this segment appears as a power-law distribution. As the evaluation standard improves, due to differences in similarity, people begin to selectively combine into a union. Under such a situation, the power-law distribution is destroyed, and the large group splits into several small groups, gradually form a lognormal distribution. From FIG.5 we can see that the modularity of this region is a growing trend, there are still a large number of structure groups in the network, so the score of the fitting function will fluctuate at this time. When the similarity threshold exceeds the phase transition point ($\theta$=0.35), modularity declines and stabilizes, the degree distribution also presents a stable lognormal distribution.

**4.3 Experiment Results of State Evolution**

**4.3.1 Analysis of the evolution results of node states**

Corresponding to the evolution of network structure, we conducted experiments on node state evolution. FIG.7 shows the evolution of the state similarity matrix among the two sets of network nodes under different iteration steps.

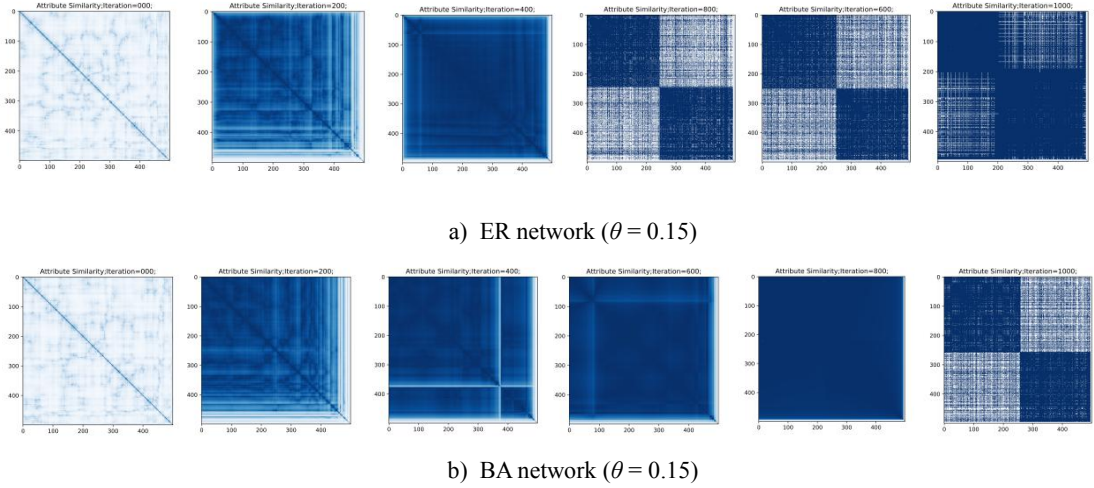

a) ER network ($\theta$ = 0.15)

b) BA network ($\theta$ = 0.15)

FIG.7 Evolution results of the state similarity matrix. The similarity matrix element $s_{ij}$ calculates the state similarity between all nodes based on Eq.2 and uses the color to represent the size of $s_{ij}$. The figure shows the state similarity matrix of the two networks when the number of iterations is 0, 200, 400, 600, 800, and 1000.

In general, as the model evolves, these networks will generate several large state groups, which will diverge as the model evolves. A large-scale group will differentiate into multiple state groups within the group. Specifically, the evolution law of the ER network is shown in FIG.7 (a). When the number of iteration steps is close to 400, a very large group of similar states has been formed, that is, large-scale consensus. But it differentiates when the number of iterative steps is around 600, indicating that this large-scale consensus will undergo internal differentiation. With further evolution, the network in the final state of the model will form a large-scale state group with similar overall states but local differences. It illustrates that the homogeneity of individual states promotes the formation of large-scale state groups, but the heterogeneity of individual structures leads to overlap states within groups. The evolution law of the BA network is shown in FIG.7 (b). The evolution of the model (it ≤200) will promote the merger of a large-scale state group. With further evolution, the states of the nodes in the BA network will be more consistent, but there will be more state overlap within the group. As the number of iteration steps continue increases, node states is getting more and more similar, most of the nodes (> 95%) in the network show the same state when the number of iteration steps is close to 800. But it will eventually differentiate into multiple state groups. Based on social identity theory, this law reflects that after the formation of consensus, individual differences remain within the group, and consensus does not eliminate individual heterogeneity.

**4.3.2 Impact of similarity threshold on node state**

Corresponding to the evolution of the structure group, the impact of the similarity threshold on node state has also been investigated. The $S_{max}/N$ values of the two networks under different similarity thresholds are shown in FIG.8. Results are the average of 30 independent experiments.

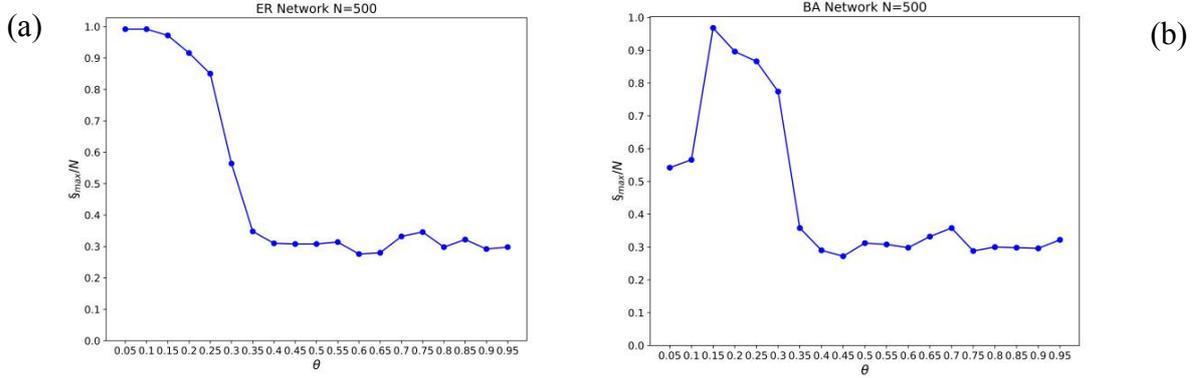

FIG.8 Distribution of $S_{max}/N$ under different similarity thresholds. The horizontal axis is the value of the similarity threshold, and the vertical axis is the $S_{max}/N$ value of the network at the last iteration.

As shown in FIG.8, the similarity threshold has a significant impact on the evolution of node state and causes a phase change in the $S_{max}/N$ value. It can be found that the phase change point is concentrated near [0.15, 0.35]. In general, as the similarity threshold increases, the $S_{max}/N$ values of the two networks tend to be stable after large fluctuations. FIG.8 (a) shows the state evolution of the ER network. With the increase of the similarity threshold, $S_{max}$ shows the change characteristics of stable and then decreasing and gradually stabilizing. When the similarity threshold interval is [0.15, 0.35], the size of the largest state group decreases from 0.99 to around 0.33. And with the increase of similarity threshold, the ratio is stabilized at around 0.3 after $\theta \geq 0.35$. It indicates that the initial network has a greater impact on the model when $\theta$ is very small. For the ER network, the similarity threshold $\theta$ has a significant impact on the state group at the interval of (0.15, 0.35). Influenced by the ER network structure, the size and structure of the state group are stable at other intervals. FIG.8 (b) shows the state evolution of the BA network. The largest state population size $S_{max}$ in the network stabilizes first and has a sudden rise, then slumps and gradually stabilizes with the increase of similarity threshold. When the similarity threshold $\theta$ is at the interval of [0.10, 0.15], the size of the largest state group increases from 0.57 to around 0.97. With the continuous increase of similarity threshold, $S_{max}$ tends to decrease at the interval of [0.15, 0.35] and stabilizes around 0.3 after $\theta \geq 0.35$. For the BA network, as similarity threshold increases, node state in the network also converges and reaches the highest point at $\theta=0.15$. However, a higher similarity threshold does not lead to convergence of node states, and only 30% of nodes are stable and will form a consistent state.

Based on the results of the above two networks, it can be found that the evolution law is generally consistent. All networks have extreme values at the interval of [0.15, 0.35], which means there exist obvious phase changes in this interval. It can be known from Eq.(6) that node state is determined by the average state of its local homogeneous neighbor nodes. And the number of homogeneous neighbor nodes is determined by Eq.(4). Therefore, the similarity threshold also affects the node state. Similar to the experimental result of the structure group, a higher or lower similarity threshold does not cause nodes to converge on the state, there also exists a critical value to cause state convergence. From the perspective of social identity theory, it can be found that the evaluation criteria of individual differences will affect the scope of consensus if we take consensus as a state of social individuals. And there exist clear critical points in the evaluation criteria of differences, which may lead to the formation of large-scale consensus. From the description of positive heterogeneity in social identity, the formation of large-scale consensus may lead to identity heterogeneity among different groups. When heterogeneity does not have positive characteristics, such as prejudice discrimination, popular and singular style, etc., groups may adopt behavior strategies to form a consensus. In conclusion, the evaluation criteria of individual differences can be used as a risk indicator to estimate the occurrence of consensus.

In summary, the similarity threshold in social media can be considered as an individual's tolerance for other people's opinions. The overall environmental tolerance in the virtual space is different from that of reality. Users can decide to follow or unfollow another user at any time, which makes the evolution of individual relationships more convenient. Individuals are more likely to form groups and exchange opinions within the group. From the existing research[1], we can know that the interaction between users in today's social media platforms such as Twitter, Facebook is very unequal. In the real world, the characteristics of relationships between celebrities and fans are similar to that of the scale-free structure in the BA network. From the results of our experiments, we can know that the opinion polarization that currently appears in social media is essentially the evolution result of the network structure. Our experimental results verify this argument, and we find that the similarity threshold has a strong control effect on this process.

For the network structure in the above process, it may evolve into a structure with lognormal distribution. This shows that under the control of the similarity threshold, a number of parallel community structures appear among polarized groups, and this community structure changes the distribution of network degrees. Cite an instance, celebrities feed cultural products to people, and people come together as fanatics community to negotiate against celebrities. This kind of tight organization brings together individuals who were unrelated before through an idol. Because they can form a community, the fan group's voice and status have been significantly improved. They are no longer ordinary audiences who were unable to speak in the dual relationship of "idol-followers", but become elders who can actively participate in the career planning of idols. In other words, the unequal power-law distribution in the original social media has changed after the emergence of the community, and it has become a lognormal distribution, which is also easy to triggered group isolation. Therefore, judging the evolution trend of the network degree distribution has certain guiding significance.

For the state distribution characteristics, it can be seen that the structure and state show different aggregation characteristics under different similarity thresholds. Generally speaking, when the threshold is around 0.15, there are obvious state groups in the network, and when the threshold is near 0.35, there are obvious structure groups. If there is a possibility that the similarity threshold may change dynamically, it indicates the inheritance relationship between the state group and the structure group. In social media, this means that if an individual's tolerance for other people's opinions slowly increases from a very low value, state groups will appear first and then structure groups.

# 5 . Conclusions

In this paper, an adaptive network model is constructed based on social identity theory. The formation of a structure group and state group has been analyzed. And we discussed the possible mechanism of formation and development of consensus in complex networks. It can be found that whether in the structure group or state group, the similarity evaluation criteria (similarity threshold) determine the final evolution result of the network. And the absolute high similarity or low similarity does not lead to differences between groups. On the contrary, when individuals tend to adjust their social relationship or state, there is often a clear threshold. Near this critical value, the network structure will be modularized or socialized, and individual states will form an obvious consensus. It indicates that the formation of consensus is often accompanied by differentiation or isolation in social groups. It also shows that the formation and development of consensus may be affected by the similarity between individuals. Moreover, through the evolution of the similarity matrix, we found that after the emergence of several large state groups, a larger-scale consensus will be formed, and then disagreement will occur, which will lead to re-differentiated in the large group. We also found that affected by the similarity threshold, the degree distribution of the BA network will change from a power-law distribution to a lognormal distribution. In summary, our model realized the adaptation of network structure and node state by restoring the social identity process of individuals. The individual similarity may determine in which way individuals interact with groups and evolutionary direction of consensus. In the real social environment, individuals' similarity belongs to a cognitive category, and its cognitive results can often be controlled through information disclosure, policy guidance, and behavioral intervention. In turn, it realizes the governance and regulation of public opinion and demonstrations.

This study is still far from completion, our model simplified the general process of an individual's pursuit of consensus because of the complexity of individual behavior. And we did not consider the direction of interaction between social individuals and the sequence of individuals updating their states. And that will be supplemented and improved in the subsequent work.

## Data Availability

The data used to support the findings of this study are available from the corresponding author upon request.

## Conflicts of Interest

The authors declare that they have no conflicts of interest.

## Acknowledgments

This work was supported by the Natural Science Foundation, Shaanxi [grant numbers 2019JQ-531]; the Social Science Foundation, Shaanxi [grant number 2019S044]; and the Funds for basic scientific research business expenses of central universities [grant number 300102238103].## REFERENCES

[1]  KWAK H, LEE C, PARK H et al. What is twitter, a social network or a news media?[C]. Proceedings of the 19th international conference on World wide web. 2010.
[2]  EVANS T, FU F. Opinion formation on dynamic networks: Identifying conditions for the emergence of partisan echo chambers[J]. Royal Society Open Science, 2018, 5(10).
[3]  FU F, WANG L. Coevolutionary dynamics of opinions and networks: From diversity to uniformity[J]. Physical